\documentstyle[12pt,aasms4]{article}

\newcommand{\vsini}{\mbox{$v\,\!\sin\,\!i$}}

\newcommand{\ha}{\mbox{H$\alpha$}}
\newcommand{\bvz}{\mbox{$(B-V)_0$}}
\newcommand{\kms}{km\ s$^{-1}$}

\newcommand{\RH}{$R_{\rm H\alpha}$}
\newcommand{\RIRT}{$R_{\rm 8498}$}

\doublespace

\begin{document}

\title{ROTATIONAL STUDIES OF LATE-TYPE STARS. VII. M34 (NGC 1039) AND
THE EVOLUTION OF ANGULAR MOMENTUM AND ACTIVITY IN YOUNG SOLAR-TYPE STARS}

\author{David R. Soderblom}
\affil{Space Telescope Science Institute\\
3700 San Martin Drive, Baltimore MD 21218\\
email: soderblom@stsci.edu}

\author{Burton F. Jones}
\affil{University of California Observatories/Lick Observatory\\
and Board of Studies in Astronomy and Astrophysics,\\ University of
California, Santa Cruz 95064\\
email: jones@ucolick.org}
\and
\author{Debra Fischer}
\affil{Department of Astronomy\\
University of California, Berkeley CA 94720\\
email: fischer@serpens.berkeley.edu}

\begin{abstract}
We analyze Keck Hires observations of rotation and activity in F, G, and
K dwarf members of the open cluster M34 (NGC 1039), which is 250 Myr old,
and we compare them to the Pleiades, Hyades, and NGC 6475.
The upper bound to rotation seen in M34 is about a factor of two lower than
for the 100 Myr-old Pleiades, but most M34 stars are well below this
upper bound, and it is the overall convergence in rotation rates that is most striking.  A few K dwarfs in M34 are still rapid rotators, suggesting that
they have undergone core-envelope decoupling, followed by
replenishment of surface angular momentum from an internal reservoir.
Our comparison of rotation in these clusters indicates that the time scale
for the coupling of the envelope to the core must be close to 100 Myr if
decoupling does, in fact, occur.

\keywords{clusters: open --- stars:late-type --- stars:evolution\\
 --- stars:rotation}

\end{abstract}

\section{Introduction}

M34 (NGC 1039) is an open cluster that is about 250 Myr old (\cite{MMM93}
1993; \cite{Jon96} 1996).  That makes it ideal for comparing to the Pleiades
(age $\sim100$ Myr: \cite{Pat78} 1978; \cite{MMM93} 1993; \cite{BMG96} 1996)
and Hyades (600 Myr: \cite{Pat78} 1978; \cite{TSL97} 1997).
Such a comparison is especially interesting for solar-type stars, which
undergo much change over that age interval.  Solar-type stars (by which we
mean roughly F7V to K2V) reach the Zero-Age Main Sequence (ZAMS) rotating
at a variety of rates, from roughly solar on up to $\sim100$ times solar,
as seen in the Pleiades (\cite{SH87} 1987; {\cite{sshj} 1993, hereafter
Paper 4), for example.  Yet almost no spread in rotation exists among the
solar-type stars of the Hyades (\cite{RTL87} 1987; Paper 4).
Chromospheric emission (CE), being intimately related to rotation,
exhibits similar differences between these clusters (Paper 4).

The differences in rotation seen between the Pleiades and Hyades are
believed to be an evolutionary effect.  The paradigm is that rotation
and convection interact in stars like the Sun to generate a magnetic
field.  One manifestation of those fields is CE\@.  The magnetic field can
grip
an ionized stellar wind well beyond the surface
of the star, leading to angular momentum loss.  In this picture, a
rapidly rotating star generates a stronger magnetic field than a slowly
rotating star, leading to a stronger field, more CE, and more rapid
angular momentum (AM)
loss.  Thus rotation rates among a coeval population may start out
very different, but should converge with time.

However, there is more than one way for a fast-rotating star in the
Pleiades to end up looking like a slowly-rotating star in the Hyades.
For example, we can suppose that
stars always rotate as solid bodies, forcing the entire
star to spin down as AM is lost.
An alternative hypothesis is that ZAMS stars may rotate at different
rates in their convective envelopes and radiative cores (\cite{SH87} 1987).
The envelope has a relatively large radius but little mass, so spinning
it down takes less time than it would to decelerate the entire star.
But the core contains most of the AM and acts as
a reservoir that can replenish the surface rotation
for an extended time.  In this case the surface rotation rate (and the
spread in rates among a group of stars) will depend
on the AM loss rate of the envelope, as well as the coupling time scale
between the core and envelope.
A coupling timescale that is short compared to the timescale for envelope AM
loss is essentially the same as
solid-body rotation.

Models for these two scenarios have been put forth by \cite{JCC93} (1993)
and
\cite{CCJ94} (1994), and the essential difference between the two cases is
as just described:  If solid-body rotation pertains, then all the stars in
the Pleiades will lose AM, and a sample of such stars will be seen to
gradually converge in its distribution of angular velocity ($\Omega$).
But if the core and envelope decouple, then the surface rates of some stars
can stay high for an extended period as they get replenished from the core,
followed by overall spindown once solid-body rotation is established.
In other words, core-envelope decoupling delays the start of the solid-body
phase.  \cite{BSP01} (2001) show that an additional factor is needed to
explain the existence of slowly-rotating stars in clusters as young as the
Pleiades, namely the time scale for coupling of the star to a disk in the
pre-main sequence phase.  This adds another parameter to the models, but
without
it the distributions of rotation seen cannot be explained, with or without
core-envelope decoupling.

It may be possible to distinguish between these models with and without
decoupling by looking at
a cluster intermediate in age to the Pleiades and Hyades, so that we
can see if such stars have changed their distributions of rotation
rates.  Such an effort was made in Paper 5 of this series
(\cite{paper5} 1993),
where rotation among stars of the Ursa Major Group was studied.
After carefully considering membership for that kinematic group,
there were not enough stars left for a
meaningful comparison.  More recently, \cite{JJ97} (1997) obtained
rotation rates for some stars in NGC 6475, a cluster that is about 220
Myr old.  Their results, shown below, reveal some fast rotators at that age,
but their sample is too small, especially for G and K dwarfs, to
provide a useful comparison to the Pleiades, and, as they note, their sample
was x-ray selected and is therefore biased toward rapid rotators.

M34, at an age of 250 Myr, is well-suited for such a study of rotation
in young solar-type stars.  It is fairly
rich, and it is close enough to make its solar-type stars accessible to
high-resolution spectroscopy with the Keck telescope and its HiRes
spectrograph.  This paper will illustrate how rotation and
CE behave among a 250 Myr old group of stars, and what that may mean
for AM evolution among stars like the Sun.

\section{Observations and Analysis}

We began this program at the Lick 3m Shane telescope, using the Hamilton
spectrograph (\cite{Vog87} 1987), which records echelle spectra on a TI
$800\times800$ CCD at a resolving power of about 50,000.  We attempted to
repeat in M34 what we had done in the Pleiades (Paper 4) for G dwarfs, but
we
had consistently bad luck with observing conditions.  The result was low
signal-to-noise spectra of some of the earlier spectral types we are
interested in, mostly mid- to late-F dwarfs.  These spectra were reduced
in the same way as the Hamilton spectra described in Paper 4.

Our primary interest, however, is in the G and K dwarfs of M34, for it is
among those stars that we hoped to find some UFRs (Ultra-Fast Rotators),
i.e., stars with $v \sin i \ga 20$ km s$^{-1}$.  Such stars are found in
the Pleiades and $\alpha$ Persei clusters, but not in the Hyades.  Just
finding a few of them in M34 would constrain AM loss models.  Also, we
wished to study CE by examining H$\alpha$ and the \ion{Ca}{2} infrared
triplet (IRT), meaning we required at least moderate signal-to-noise.

This led us to observe M34 members with the HiRes spectrograph (\cite{Vog92}
1992) on the Keck telescope.  Targets were chosen from the membership study
of
\cite{Jon96} (1996), and range in $(B-V)$ from 0.44 to 1.29.
The HiRes spectra included H$\alpha$, the Li $\lambda6708$ feature (which is
reported on in \cite{JFS97} 1997), and two of the three IRT lines
($\lambda8498$ and $\lambda8662$).

The detector for the Keck observations was a Tektronix $2048\times2048$ CCD
having 24 $\mu$m pixels.  We used the ``C1'' slit decker, which yielded a
projected slit height of 7 arcsec (36.6 pixels), and a projected slit width
of 0.86 arcsec (3 pixels), for a resolving power of 45,000.  The wavelength
range was about 6300 to 8700 \AA, divided into 16 orders, with each order
covering about 100 \AA.  This means the order coverage is incomplete.
Integration times ranged from 5 to 40 minutes, and that gave a typical
signal-to-noise of 70.

We obtained rotation (\vsini) and radial velocity ($v_{\rm rad}$) measures
by cross-correlating the spectra against narrow-lined templates, choosing
a template with a temperature similar to the program star.  Standard star
spectra were also cross-correlated with spun-up templates, and the FWHM was
measured.  These procedures are described in \cite{JFS97} (1997).  The
emission fluxes in the H$\alpha$ and the 8498 \AA\ line of the IRT were
determined in the same way as in Paper 4, namely by first subtracting the
profile of an inactive star of the same intrinsic color (see Table 3 of
Paper 4), measuring the equivalent width of the resultant emission
feature, and then converting that $W_\lambda$ to a flux.  Finally, the
fluxes were ratioed to the stellar bolometric flux to calculate \RH\ and
\RIRT.  The appropriate formulae are given in Paper 4.
Our results are listed in Table 1.
\marginpar{Tab. 1}
Some values in Table 1 differ from what was reported in \cite{JFS97} (1997).
First, in that paper we formally derived \vsini\ values as low as 5 \kms,
but here we list any value below 7 \kms\ as ``$\le7$'' to be more
conservative.  Second, for three stars we could derive \vsini\ values for
both components of spectroscopic binaries; these are JP 156, 257, and 317.

\section{Discussion}

\subsection{Angular Momentum Loss After the ZAMS}

The essence of our observations of rotation in M34 stars is illustrated
in Figure 1.
\marginpar{Fig. 1}
The first panel shows rotation in the Pleiades (Paper 4), to which two
lines have been added.  The upper curve represents the upper
bound of the Pleiades distribution of rotation, while the lower curve is
the mean of rotation in the Hyades (panel d, from Paper 4).
One key conclusion
of Paper 4 was that the rotation of solar-type stars, in evolving from the
Pleiades to the Hyades, changes only modestly in the mean, but undergoes
a huge convergence in rotation rates.  Thus, at any one mass in the
Pleiades, the rotation rates differ by an order of magnitude or more,
yet in the Hyades stars of the same mass have nearly identical rotation
rates.  This effect is shown here in \vsini, but it has been confirmed
through measurements of rotation periods (\cite{RTL87} 1987; \cite{PSS93}
1993; \cite{PSM93} 1993; \cite{KT98} 1998).

Figure 1(b) shows the \vsini\ observations of \cite{JJ97} (1997) for NGC 6475
(220 Myr old).   The presence of two stars with \vsini$\approx20$ \kms\
at \bvz$\approx1.1$ suggests that rapid rotators exist at that age,
but the small sample size precludes drawing more conclusions.

Figure 1(c) shows our results for M34.  For the bluer stars
[\bvz$\la0.7$], the rotation rates are similar to most Pleiads in this
color range, except that the upper branch of the Pleiades rotational
distribution (stars with \vsini$\approx50$ \kms) has disappeared.
For the stars redder than 0.7 in \bvz, few if any members of M34 have spun
down to the low rates seen in the Hyades (represented by the lower line
in each plot and shown in Fig. 1(d)).  Many stars have \vsini\ values of
about 8 to 15 \kms,
essentially the same as seen in the Pleiades, but the upper envelope
for M34 is about a factor of two lower than for the Pleiades (as
represented by the upper line).  Also, the proportion of rapid rotators
in M34 (those with \vsini$\ga15$ \kms) is the same as the proportion
of UFRs (stars with \vsini$\ga30$ \kms) seen in the Pleiades, given
the sample sizes.  Finally, M34 contains at least one blue UFR (JP 158,
at \bvz=0.80 and \vsini=45 \kms).

It is clear that the convergence in rotation rates exceeds any overall
decrease in the average rotation rate as solar-type stars evolve from
the age of the Pleiades to the age of M34 and on to the age of the Hyades.
Stars that rotate relatively
slowly at the age of the Pleiades will still be rotating at about the
same rate much later.  For example, \cite{KT98} (1998) see Pleiades
stars with rotation periods of 7 to 8 days at \bvz$\approx1.1$, while at
that same color \cite{RTL87} (1987) observe rotation periods of 11 to 12
days in the Hyades, a change by a factor of about 1.5;
our M34 \vsini\ values for the slow rotators are consistent with this trend.
Yet in that same 500 Myr, the surface rotation rates of the
UFRs have declined by a factor of twenty or more.

The calculations done so far to study AM loss in young solar-type stars
have, of necessity, adopted the Pleiades and Hyades as being fully
representative of any ensembles of stars of their ages.  Yet the Pleiades
and Hyades differ in one important respect (besides age): their
metallicities.  The Pleiades has near-solar abundances ([Fe/H] =
$+0.06\pm0.05$; \cite{KS2000} 2000),
but the higher metallicity of the Hyades ([Fe/H] = $+0.127\pm0.022$;
\cite{BF90} 1990),
means that Hyads have deeper convective envelopes at a given mass.
Other things being equal, this may lead to stars in the Hyades losing
AM more quickly than solar-metallicity stars do.  The importance of
this can be gauged by pretending that a given Hyad acts like a
solar-metallicity star of slightly lower mass.  A comparison of the
AM loss tracks of \cite{CCJ94} (1994) and the models of \cite{SPT00}
(2000) for different masses show
that lower-mass stars indeed lose AM more quickly, but the net difference
is modest.  In other words, there appears to be no valid reason to reject
the Hyades as a basis of comparison for the observations.

\subsection{Activity in M 34}

Figure 2
\marginpar{Fig. 2}
shows \RH\ versus \bv\ for stars in the Pleiades (from Paper 4),
NGC 6475 (\cite{JJ97} 1997), and M34.  Among the M34 stars we observed,
there are five that exhibit ``overt'' \ha\ emission; i.e., portions of
the line rise above the continuum.  These have been highlighted in Fig. 2(c)
with overlaid crosses, and their \ha\ profiles are shown in Figure 3.
\marginpar{Fig. 3}
In Paper 4 we also showed six examples of Pleiades
stars with overt \ha\ emission, and those shown in Fig. 2(a)
have also been highlighted.

\ha\ emission in M34 is similar to that seen in the Pleiades.  All
the M34 stars with \vsini $>15$ \kms\ and \bv$\ge0.80$ show overt \ha\
emission, although few Pleiads above the same thresholds do.  The lack
of Pleiads with overt emission is probably not simply due to rotational
broadening of the lines.  This can be seen by noting that in the same
color range of $(B-V)\approx1.0$ to 1.1 there are M34 stars with overt
emission having \vsini\ as great as Pleiads that lack overt emission.

Figure 4
\marginpar{Fig. 4}
compares the Pleiades and M34 with regard to the relation between \ha\
emission and a pseudo-Rossby number, $N^\prime_{\rm R}$.
The Rossby number is the ratio of a
star's rotation period to its convective turnover time, the latter being
calculated from models (see Paper 4).  Only an upper limit to $P_{\rm rot}$
can be determined from \vsini, of course, but that is sufficient here.
The details of the calculation are given in Paper 4, eq. (7).

At a given value of $N^\prime_{\rm R}$, stars in M34 and the Pleiades
have comparable levels of \ha\ emission overall.  However, any star can
be moved to the right in this diagram because the measured \vsini\ may not
show the full extent of the true rotation ($v_{\rm eq}$).  Within a
cluster, those stars with the largest \vsini\ values are probably seen
equator-on, and so it is more likely that Pleiades stars with
$\log N^\prime_{\rm R}$ from $-1.0$ to $-1.4$ in fact have true
$\log N^\prime_{\rm R}$ values that are smaller (i.e., approaching $-2$)
because those stars do not have the largest \vsini\ values seen in their
color range.

Figure 5
\marginpar{Fig. 5}
shows a comparison of \RH\ to \RIRT\ for M34 stars.  As for the Pleiades
(Paper 4), different chromospheric indices yield different pictures of
the chromosphere, with scatter that exceeds observational error, but
there is a broad correlation between the quantities.

One expects the stars with overt \ha\ to also emit strongly in x-rays.
\cite{Sim2000} (2000) has recently published Rosat observations of M34, and
of the five stars that emit in \ha, only two were detected by him.  Of the
remaining three, two (stars 265 and 356) would have been at the very edges
of
the Rosat field (where the sensitivity is lower),
but star 536 is only 12 arcmin from the center of the field and
so should have been detected.  The small number of stars in common between
our
sample and Simon's prevents firm conclusions from being drawn, but it
appears possible that not all strong \ha\ emitters are also strong in x-rays.

\section{Core-Envelope Decoupling versus Solid-Body Rotation}

It has been hypothesized that the surface AM of some stars is steadily
replenished by an internal reservoir over
time when they are young.  Can we entirely rule out solid-body rotation?
\cite{CCJ94} (1994) show calculations of the evolution of rotation when no
decoupling occurs.  In their Figure 3 the net change in $\Omega$ (the
angular
velocity) for the slow rotators in going from 100 to 250 Myr is only about
0.1 dex, an amount that may be undetectable in our observations of \vsini.
The rapid
rotators in their models drop by about 0.3 dex in $\Omega$ over this time
interval, just as we see them do.  For the slow rotators, the change in the
models going from 100 Myr to 600 Myr is 0.3 dex, or a factor of two.  The
actual change we see is comparable to this.

If core-envelope decoupling occurs, it effects should be most evident at
an intermediate age like that of M 34 among stars that started as UFRs.  By
comparison, the slow rotators are probably spinning as solid bodies for the most part, and the very fastest rotators in the Pleiades must also be solid bodies
because they already have extremely short periods (as little as 0.25 day): how
much faster could a core be spinning?  In M 34 we see intermediate-level
rotation rates at \bv\ near 1.0, while the more massive stars have mostly converged to similar rates.  Why do these K dwarfs still show appreciable
surface rates?  Either they are inherently less efficient at losing AM,
which is contradicted by models, or they have tapped into an internal
reservoir.
It may be possible to
construct models with solid-body rotation that match the observations
better, but the easiest way to account for all the data is to
invoke core-envelope decoupling.
As we have pointed out before (Paper 4), core-envelope decoupling
also helps to account for the strong mass dependence of rotation in
solar-type stars by the way in which the ratio of the moment of inertia
of the core to that of the envelope vary with mass.
Observations of rotation periods for a significant sample of stars in a
cluster about 200 to 300 Myr old will, we believe, settle this issue by
showing just how slowly the slowest-rotating stars spin.

\cite{BSP01} (2001) also favor core-envelope decoupling in their models of
disk locking in the early evolution of solar-type stars.  They note that
models with solid-body rotation require disk locking for 10 to 20 Myr.
This is not impossible, but it is uncomfortably long when other observations
suggest lifetimes for optically thick disks that are well
under 10 Myr (\cite{HYJ00} (2000).

For the moment,
assume that core-envelope decoupling indeed takes place in order to
see what constraints we can place on the time scale for angular momentum
transfer ($\tau_{\rm coupl}$) between the surface and the inner regions.
First, we note that $\tau_{\rm coupl}$ cannot be much less than about
100 Myr, or otherwise core-envelope decoupling would never occur in the
first place.  This is because the PMS evolutionary time scales for
low-mass stars can exceed 100 Myr, yet we see UFRs in those masses.
At the same time, the near-uniformity of rotation rates seen in the
Hyades, at an age of 600 Myr, suggests that several $\tau_{\rm coupl}$
have occurred in the 500 Myr from the Pleiades to the Hyades.  If
$\tau_{\rm coupl}$ were as much as 500-1000 Myr, then Hyades stars could not
have achieved
such uniform rotation rates, and if $\tau_{\rm coupl}\sim2$ Gyr, then the
Sun would not show solid-body rotation (unless it started as a slow rotator).
All this means
that $\tau_{\rm coupl}$ is unlikely to exceed 100 Myr by much, and so, to first
order, if decoupling occurs then the recoupling timescale,
$\tau_{\rm coupl}$, is $\sim100$ Myr.

\acknowledgments
These observations were made in part at the W. M. Keck Observatory.
The W. M. Keck Observatory is operated as a scientific partnership between
the California Institute of Technology and the University of California.
It was made possible by the generous financial support of the W. M. Keck
foundation.  D.S. acknowledges support from NASA's Ultraviolet, Visible, and
Gravitational Physics Research and Analysis Program.

\newpage

\figcaption{
Rotation rate (\vsini, in \kms) versus dereddened \bv\ for solar-type
stars in four clusters.  The upper line is meant to schematically
outline the upper bound to rotation for the Pleiades.  The lower line
is a mean relation for the Hyades from Paper 4.
(a) The Pleiades, 100 Myr old, from Paper 4.  The triangles
represent upper limits to \vsini.
(b) NGC 6475, 220 Myr old, from \cite{JJ97} (1997).
(c) M 34, at 250 Myr.
There are four stars in the Pleiades and five in M 34 that exhibit H$\alpha$
emission that rises above the continuum and their points are overlaid with
crosses.
(d) the Hyades, at 600 Myr.  For the Hyades, the open symbols
represent \vsini\ values, while the filled symbols are determined from
observations of rotation periods, and so have no aspect dependence.
}

\figcaption{
Normalized indices of \ha\ emission, \RH, versus dereddened \bv\ color
for (a) the Pleiades (from Paper 4), (b) NGC 6475 (from
\cite{JJ97} 1997), and, (c) M 34.  The vertical dotted
lines for two Pleiades stars connect maximum and minimum estimates of
\RH\ for those two stars because both show evidence of emission over
the full breadth of the profile, not just in the core of \ha\ (see Paper 4).
There are five stars in the Pleiades and five in M 34 that exhibit H$\alpha$
emission that rises above the continuum and their points are overlaid with
crosses.
}

\figcaption{
\ha\ profiles for five M34 stars with ``overt'' \ha\ emission, i.e.,
emission that rises above the continuum.
}

\figcaption{
Normalized indices of \ha\ emission, \RH, versus $N^\prime_{\rm R}$, the
normalized \vsini\ equivalent to a Rossby number.  For the Pleiades,
upper limits to \vsini\ are not shown, and for both the Pleiades and
M34 stars with overt \ha\ emission are highlighted.  In the bottom panel,
the solid points are for M34 and the crosses are for the Pleiades.
}

\figcaption{
Normalized indices of Ca {\sc ii} IRT emission, \RIRT, versus \RH\
for stars in M34.  As expected, the two quantities are well-correlated
overall, but the correlation breaks down among the weak emitters.
}

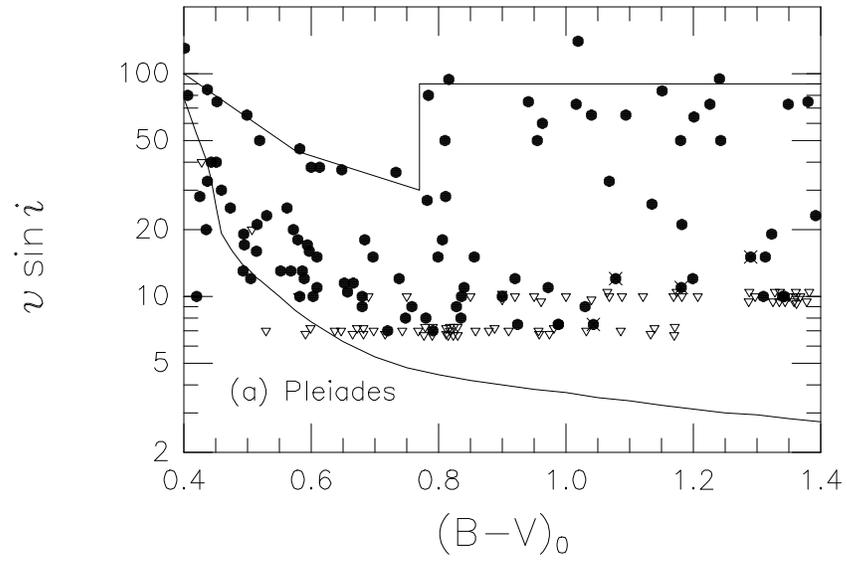
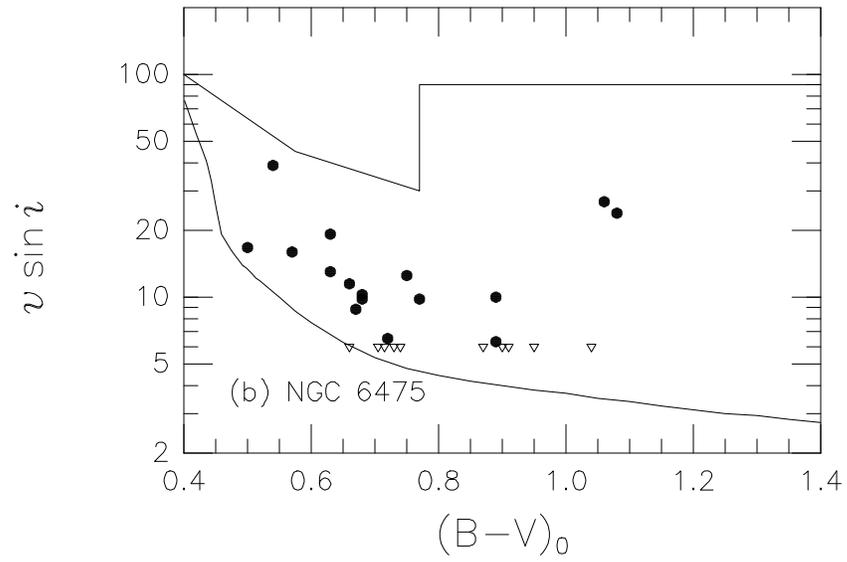
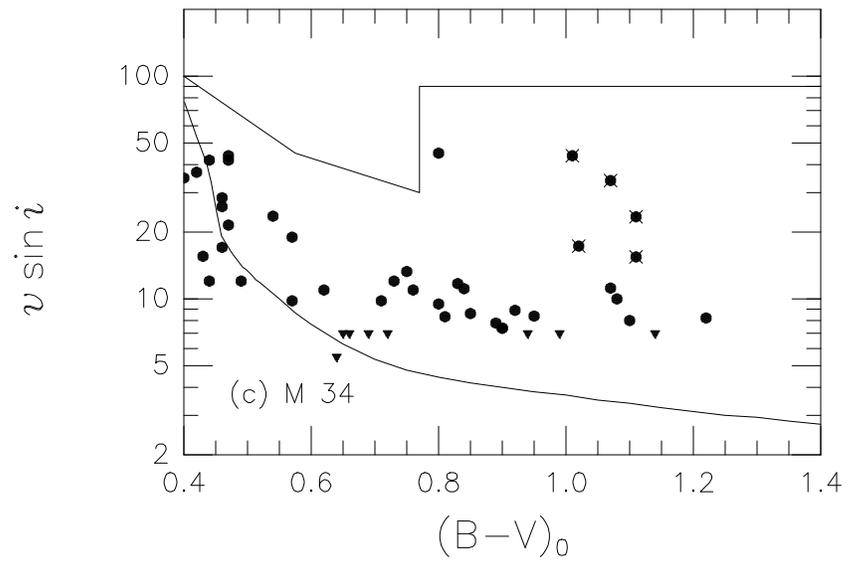
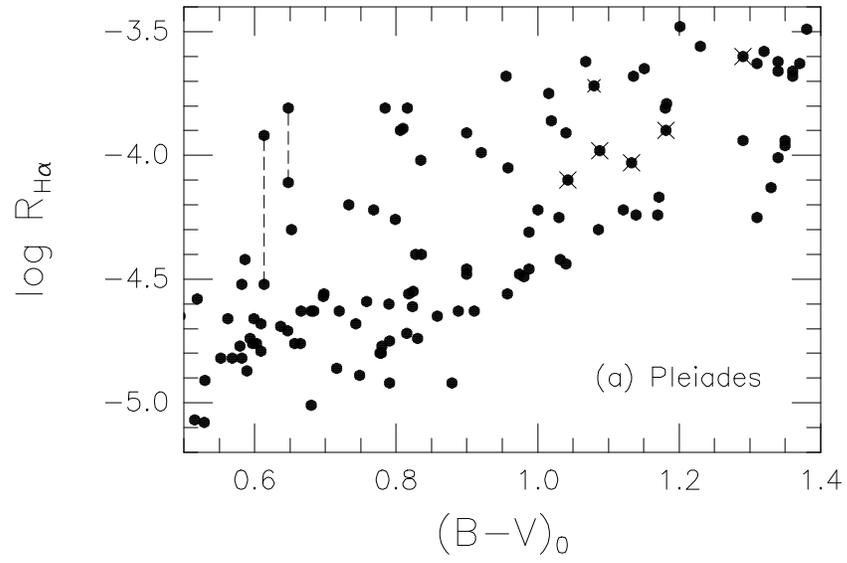
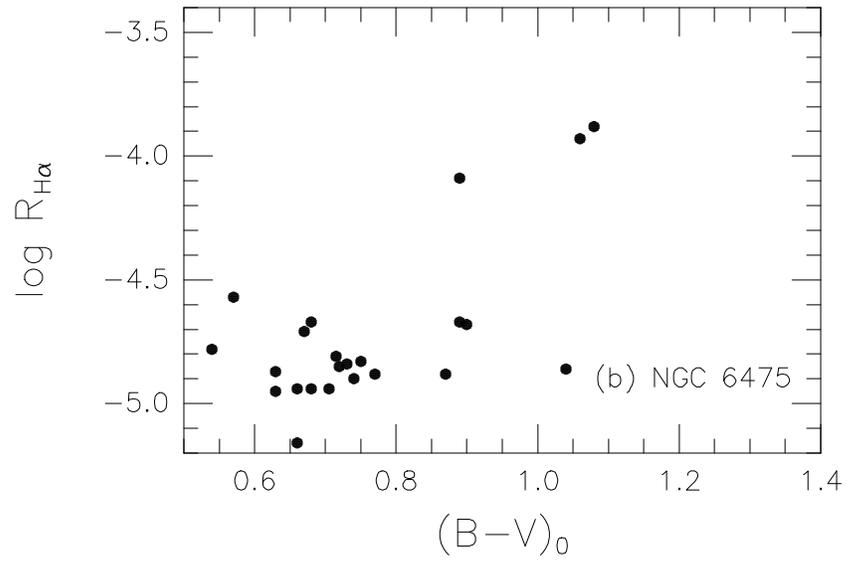
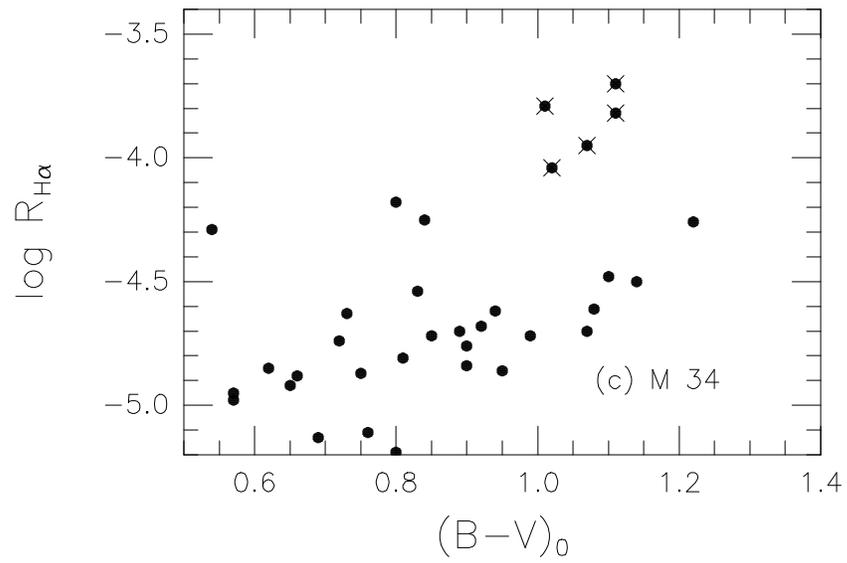
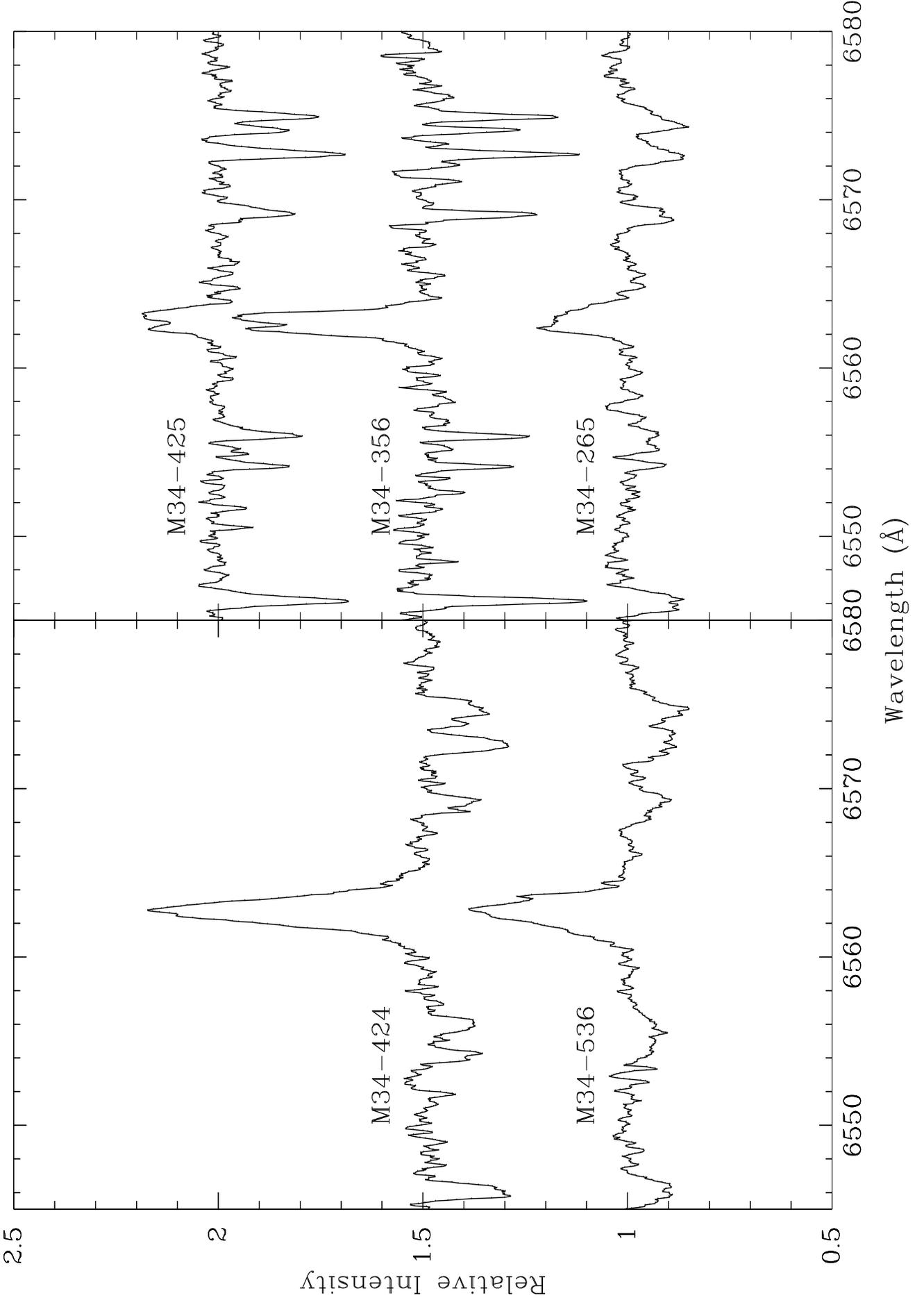
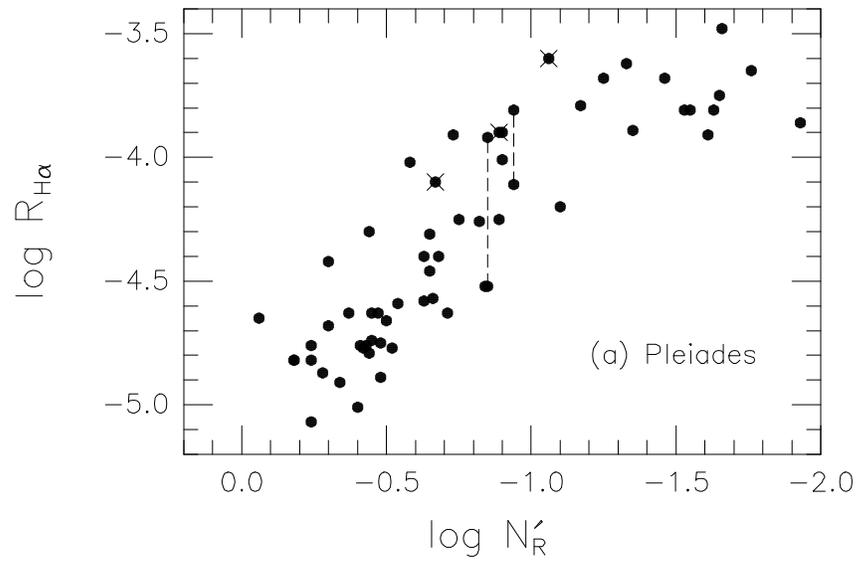
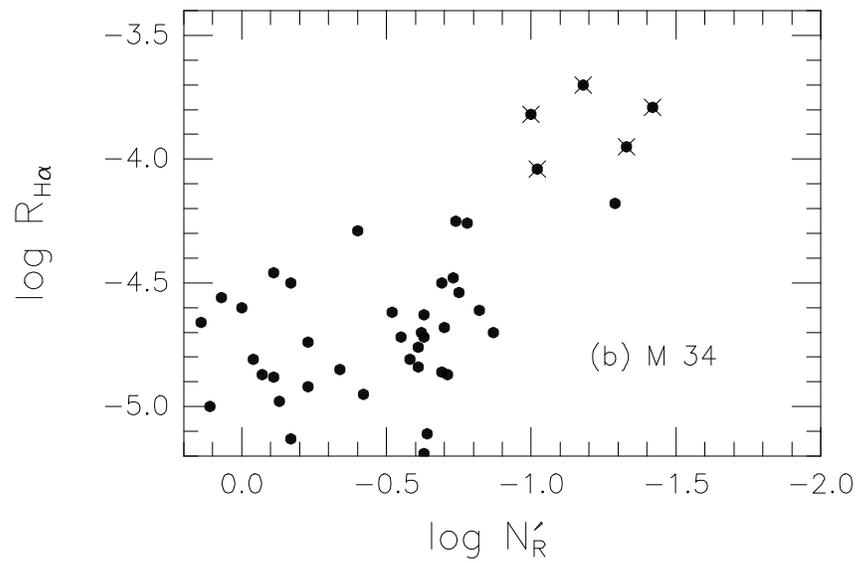
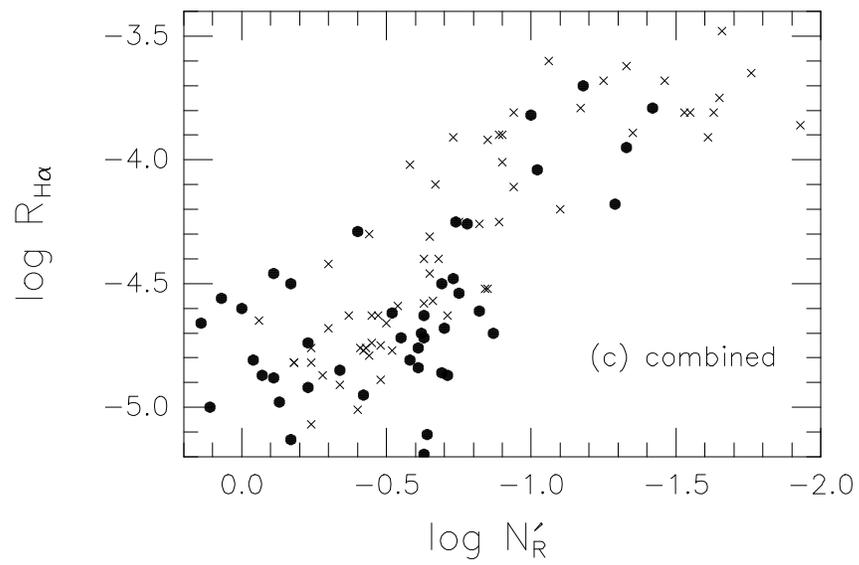
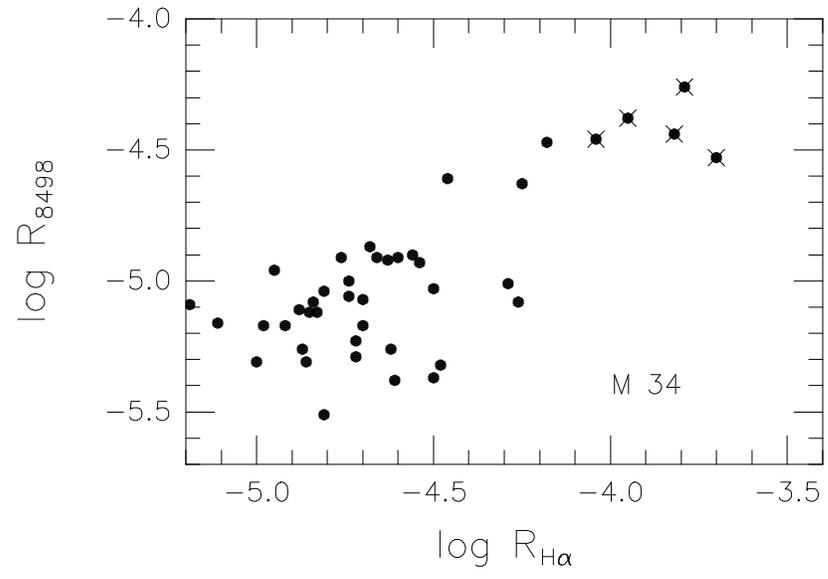
\end{document}